\documentclass[useAMS,usenatbib]{mn2e}
\usepackage{graphicx}

\usepackage{journal_names}

\title[Megahertz peaked-spectrum sources in the Bo\"{o}tes field.]{Megahertz peaked-spectrum sources in the Bo\"{o}tes field I - a route towards finding high-redshift AGN?}
\author[Rocco Coppejans et al.]{Rocco Coppejans$^{1}$\thanks{Email: r.coppejans@astro.ru.nl}, David Cseh$^{1}$, Wendy L. Williams$^{2,3}$, Sjoert van Velzen$^{4}$ and \newauthor Heino Falcke$^{1,3,5}$ \\
$^{1}$Department of Astrophysics/IMAPP, Radboud University Nijmegen, P.O. Box 9010, 6500 GL Nijmegen, The Netherlands\\
$^{2}$Leiden Observatory, Leiden University, PO Box 9513, 2300 RA, Leiden, The Netherlands\\
$^{3}$Netherlands Institute for Radio Astronomy (ASTRON), PO Box 2, 7990 AA, Dwingeloo, The Netherlands\\
$^{4}$Department of Physics and Astronomy, The Johns Hopkins University, Baltimore, MD 21218, USA\\
$^{5}$Max-Planck-Institute f\"{u}r Radioastronomie, auf dem H\"{u}gel 69, 53121 Bonn, Germany\\}

\begin{document}

\date{}

\pagerange{\pageref{firstpage}--\pageref{lastpage}} \pubyear{2014}

\maketitle

\label{firstpage}

\begin{abstract}
We present a 324.5\,MHz image of the NOAO Bo\"{o}tes field that was made using Very Large Array (VLA) P-band observations. The image has a resolution of $5.6\times5.1$\,arcsec, a radius of $2.05^\circ$ and a central noise of $\sim0.2$\,mJy\,beam$^{-1}$. Both the resolution and noise of the image are an order of magnitude better than what was previously available at this frequency and will serve as a valuable addition to the already extensive multiwavelength data that are available for this field. The final source catalogue contains 1370 sources and has a median 325 to 1400\,MHz spectral index of -0.72. Using a radio colour-colour diagram of the unresolved sources in our catalogue, we identify 33 megahertz peaked-spectrum (MPS) sources. Based on the turnover frequency linear size relation for the gigahertz peaked-spectrum (GPS) and compact steep-spectrum (CSS) sources, we expect that the MPS sources that are compact on scales of tens of milliarcseconds should be young radio loud active galactic nuclei at high ($z>2$) redshifts. Of the 33 MPS sources, we were able to determine redshifts for 24, with an average redshift of 1.3. Given that five of the sources are at $z>2$, that the four faint sources for which we could not find redshifts are likely at even higher redshifts and that we could only select sources that are compact on a scale of $\sim5$\,arcsec, there is encouraging evidence that the MPS method can be used to search for high-redshift sources.
\end{abstract}



\section{Introduction}
\label{sec:introduction}

The National Optical Astronomy Observatory (NOAO) deep wide-field survey Bo\"{o}tes field covers $\sim 9$\,$\rm deg^2$ in the optical and near infrared $\rm B_{\rm W}$, R, I and K bands \citep{ndwfs}. In addition, the field has also been covered in the X-ray \citep{Murray2005,Kenter2005}, ultraviolet \citep{Martin2003}, mid infrared \citep{Eisenhardt2004,Martin2003} and radio: 3.1\,GHz \citep{Croft2013}, 1.4\,GHz \citep{deVries2002bootes,Higdon2005}, 153\,MHz \citep{Williams2013,Intema2011} and at 62\,MHz, 46\,MHz and 34\,MHz by \citet{vanweeren2014}. Furthermore, the active galactic nuclei (AGN) and Galaxy Evolution Survey \citep[AGES;][]{Kochanek2012} have measured redshifts for 23,745 galaxies and AGN in the field. This extensive multiwavelength data set have in the past been used to study, amongst other things: the properties of galaxy groups \citep{Vajgel2014}, the spacial clustering of quasars \citep{Hickox2011}, search for radio transients \citep{Bower2011}, study the start formation rate in optical obscured galaxies \citep{Calanog2013} and study the composition of gas and dust in AGN \citep{Usman2012}. The 325\,MHz observations of the field presented here will serve as a valuable addition to the multiwavelength coverage of the field and will, as shown in this paper, be used to study these topics and more.   

Supermassive black holes, which drive AGN, are believed to lie at the center of nearly every galaxy. They not only grow with their host galaxy, but also shape and influence the galaxy and inter-galactic medium via feedback from their relativistic jets. So, to understand galaxies and their evolution we need to understand AGN evolution \citep[eg.][]{fabian2012}. A critical key to this is identifying both young AGN and AGN at high-redshifts, something that is currently a great observational challenge.

While AGN have been found in optical surveys out to $z=7.1$ \citep{mortlock2011}, Ly-alpha absorption makes detecting them at $z>6.5$ very difficult \citep{mortlock2011,becker2001}. On the other hand, the most powerful AGN in the radio, such as Cygnus A, are expected to have observed 1\,GHz flux densities in the order of 15\,mJy at $z=8$, well within the capabilities of modern radio telescopes. Hence, not only should these sources be detectable in the radio, but radio also has the advantage that it is not obscured by dust which can hide sources at all redshifts from optical identification. 

While large-scale radio surveys such as the Very Large Array (VLA) Faint Images of the Radio Sky at Twenty-Centimeters (FIRST) survey \citep{first} can detect these sources, a method needs to be found to identify them based purely on their radio properties. Based on an observed correlation between redshift and spectral index \cite[eg.][]{DeBreuck2000}, searches for high-redshift radio galaxies tend to focus on sources with ultra-steep spectral indices. While this method has proven successful in finding sources out to $z \sim 4$ \citep[eg.][]{jarvis2001,cruz2006,deBreuck2006}, \citet{singh2014} found median redshifts of only $\sim1.18$ and $\sim1.57$ for their two selections of faint ultra-steep-spectrum (USS) sources, which is only slightly higher than the median redshifts of $\sim0.99$ and $\sim0.96$ for the non-USS sources in their samples. Moreover, the physical reason for why USS sources should be at higher redshifts than non-USS sources remains unclear \citep{miley2008} and indeed recent deep observations of the Cosmological Evolution Survey (COSMOS) field with the VLA found no clear evidence that sources at higher redshift have steeper spectral indices \citep{smolcic2014}. Finally, in a recent study using nine samples of radio sources, \citet{ker2012} found that the fraction of $z>2$ sources is not significantly higher in the sub-sample of USS sources compared to the full sample. Hence, finding a method with which high-redshift AGN can reliably be selected from radio images (the focus of this paper) is very important as it would allow us to search for them in existing large radio surveys such as FIRST, current and future LOFAR \citep[Low-Frequency Array;][]{lofar} surveys such as the Multifrequency Snapshot Sky Survey (MSSS) and in the future, the Square Kilometer Array (SKA), which might be able to detect these sources beyond redshift ten \citep{falcke2004}.

High-frequency peaked (HFP), Gigahertz peaked-spectrum (GPS) and compact steep-spectrum (CSS) sources are thought to be the youngest and smallest AGN \citep{o'dea1998,Murgia2002,Conway2002}. These sources are characterized by their peaked spectra and steep spectral indices ($\alpha$; where $\alpha$ is defined as $S \sim \nu ^ \alpha$ with $S$ the spectral flux density at frequency $\nu$) above the spectral turnover. The HFP and GPS sources have rest-frame turnover frequencies above 1\,GHz and linear (lobe to lobe) sizes smaller than $\sim1$\,kpc while the CSS sources typically turn over below 500\,MHz and range in size from 1\,kpc to 20\,kpc. An evolutionary sequence is believed to exists from the HFP to the GPS sources and finally to the CSS sources \citep{o'dea1998,snellen2000,tschager2003}. Furthermore, it is likely that the CSS sources will evolve to eventually become the large FR\,I and FR\,II radio galaxies \citep{Begelman1996,snellen2000,devries2002}.

An empirical relation was found between the rest-frame turnover frequency $\nu_r$ and the linear size of nearby ($z\sim1$) GPS and CSS source over three orders of magnitude \citep{o'dea1998,snellen2000}. From the observed turnover frequency $\nu_o$ we can therefore estimate the linear size of a source by taking into account that $\nu_r=\nu_o(1+z)$, where $z$ is the redshift of the source. Using this, we can predict the source's observed angular size, noting that in conventional cosmology ($\Omega_{\rm m}=0.3$, $\Omega_{\lambda}=0.7$, $H_0=72$\,km\,s$^{-1}$\,Mpc$^{-1}$), a fixed ruler will appear to first decrease in angular size moving from $z=0$ to $z\sim1$ before appearing to increase again beyond $z\sim1$. We therefore expect a typical GPS source of 100\,pc, with corresponding rest-frame turnover frequency of 2.7\,GHz, to appear as a 25\,mas source at $z=10$ with an unusually low observed turnover frequency of 245\,MHz. From these arguments, \citet{falcke2004} proposed an, as of yet, untested method to find high-redshift AGN by searching for compact (on scales of tens of milliarcseconds) megahertz peaked-spectrum (MPS; turnover frequency below 1\,GHz) sources.

It should be pointed out that, while our aim is to use the MPS method to search for high-redshift ($z>2$) AGN, we expect that the method will also find young AGN at lower redshifts. For example, sources with a rest-frame turnover frequency of 1\,GHz and corresponding size of 470\,pc will have an observed turnover frequency of 325\,MHz and angular size of 60\,mas at $z=2.1$. While this means that we do not expect all the MPS sources that are compact on a few hundred milliarcseconds to be at high redshifts, the sources that are not at high redshifts are still young AGN and are interesting in their own right since they help us understand how the AGN population evolves with redshift. 

In this paper, we publish a 325\,MHz source catalogue\footnote{Available in electronic format from the CDS at http://cdsarc.u-strasbg.fr/} of the Bo\"{o}tes field that is at least an order of magnitude better in terms of both resolution and noise (see Table \ref{tbl:matching catalogues}) than what was previously available for this field with the Westerbork Northern Sky Survey \cite[WENSS;][]{wenss}. We also take the first steps towards using the MPS method to search for high-redshift AGN by identifying 33 MPS sources in the image and looking at their redshift distribution.

In Section \ref{sec:data reduction} we describe the data reduction and imaging before describing source extraction, image quality and catalogue construction in Section \ref{sec:matching and flux scale}. The spectral properties of the sources, selecting and discussing the MPS sources and the affect of selecting USS sources in different frequency ranges on the selected sample of USS sources is presented in Section \ref{sec: results and discussion}. Finally, in Section \ref{sec:summary} we summarize our results and discuss future work.


\section[]{Observations and Data Reduction}
\label{sec:data reduction}

The archival data set (project code AB0976) of the Bo\"{o}tes field was taken with the VLA on 22 November 2000, 12 December 2000, 2 January 2001 and 4 January 2001 and is summarized in Table \ref{tbl:observations}. The observations were taken in the P-band with the VLA in A configuration, using 27 antennae on every day except 4 January 2001 when only 26 antennae were available. The observations covered a single pointing in the Bo\"{o}tes field centered at RA 14:32:05.72 and DEC +34:16:47.5 using 3.3\,s integrations with a total integration time on the field of 20 hours spread over the four days. The data were taken in spectral line mode with two intermediate frequencies (IFs) at 327.53\,MHz and 321.53\,MHz each with two polarizations (LL and RR). We will take the average frequency of 324.525\,MHz of the two IFs as the representative frequency. Each IF was split into 16 channels of 394\,KHz each, giving a bandwidth of 6.3\,MHz per IF and total bandwidth of 12.6\,MHz. Observations of the calibrator source 3C286, used for phase, amplitude and bandpass calibration, were irregularly interleaved between those of the target field. Typically, 3C286 was observed for 2.5\,minutes every 25\,minutes.

\begin{table}
 \centering
 \begin{minipage}{\columnwidth}
  \caption{Summary of observations.}
  \begin{tabular}{ccccc}
  \hline
  Date & Start Time & End Time & TOS $^{\rm a}$ & TOS $^{\rm a}$  \\
       & [UTC]      & [UTC]    & Bo\"{o}tes     & 3C286 \\ 
       &            &          & [min]          & [min] \\
  \hline  
   22 Nov. 2000 & 12:06:59 & 20:02:56 & 255 & 42\\
   12 Dec. 2000 & 12:17:30 & 20:14:00 & 280 & 34\\
   2 Jan. 2001  & 10:22:56 & 18:51:20 & 285 & 38\\
   4 Jan. 2001  & 10:47:40 & 18:43:33 & 381 & 41\\
  \hline
  \multicolumn{5}{p{\columnwidth}}{\footnotesize{\textbf{Notes:} $^{\rm a}$ The time on source (TOS).}}\\
  \end{tabular}
  \label{tbl:observations}
 \end{minipage}
\end{table}

\begin{figure*}
  \includegraphics[width=\textwidth]{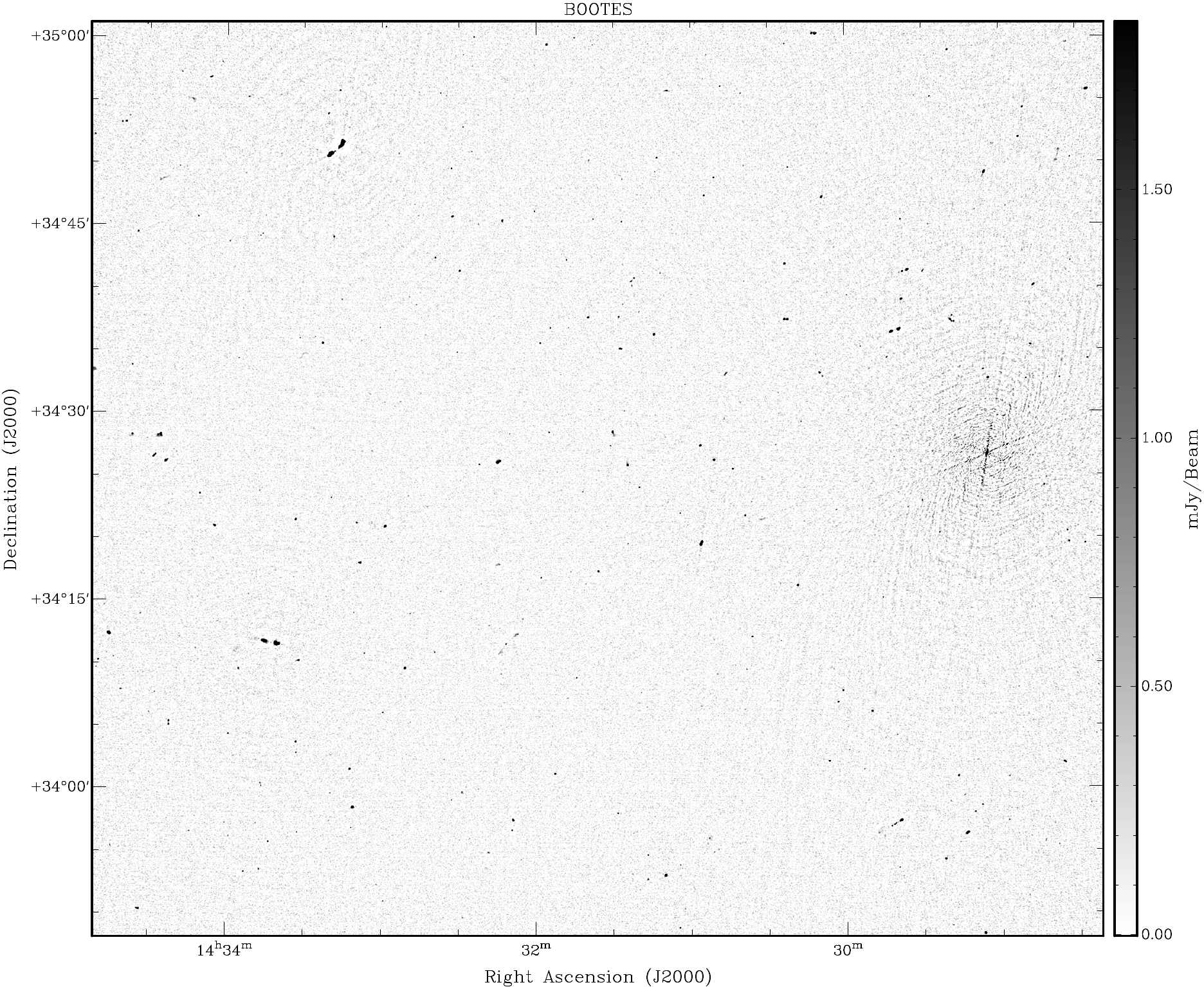}
  \caption{Inner $1.2^\circ$ of the VLA-P image. The full image has a radius of $2.05^\circ$, a resolution of $5.6\times5.1$\,arcsec and central noise of $\sim0.2$\,mJy\,beam$^{-1}$.}
  \label{fig:image_of_field}
\end{figure*}

The initial data reduction was done using the \textsc{aips} \citep{aips} software package. Data were flagged using both the automatic flagging task \textsc{rflag} and the manual flagging task \textsc{tvflg} before and after calibration. After calibration, the calibrated data were imported into \textsc{casa}\footnote{http://casa.nrao.edu} for self-calibration, imaging and primary beam correction. Five rounds of phase only self-calibration was done, in which the solution interval was decreased from 20 to 1\,min. During each round, a clean component model of the brightest sources in the field was derived from the previous best image and used as an input model. The final image of the field (the inner $1.2^\circ$ of which is shown in Fig. \ref{fig:image_of_field}) was made using 1.5\,arcsec pixels, Brigg's weighting with a robustness parameter of 0.0 and 128 w-projection planes. While the VLA P-band primary beam has a full-width half-maximum (FWHM) of $2.5^\circ$, the final image produced using \textsc{casa's clean} task was made with a diameter of $4.1^\circ$ to increase the number of sources in the image. The image has a resolution of $5.6\times5.1$\,arcsec and central noise of $\sim0.2$\,mJy\,beam$^{-1}$ increasing to $\sim0.8$\,mJy\,beam$^{-1}$ at the edge of the image after the flux correction has been applied (see Section \ref{subsec:flux scale}).

\section[]{Image Quality and Catalogue Construction}
\label{sec:matching and flux scale}

\subsection{Source Extraction}
\label{subsec:source extraction}
Source extraction was done using the \textsc{pybdsm} source detection package\footnote{http://dl.dropboxusercontent.com/u/1948170/html/index.html}. \textsc{pybdsm} works by identifying all pixels in the image that are above the pixel detection threshold and adding all of the surrounding pixels that are above the island detection threshold together to form islands of emission. Multiple Gaussians are then fitted to each island before grouping the Gaussians together into individual sources. We used the default values of three and five times the the local root mean square (rms) noise for the island and pixel thresholds respectively, where the local rms noise was calculated in a sliding box of $50\times50$ pixels that was moved across the image in steps of ten pixels. 

For each source, the flux density reported is the sum of the flux densities of the Gaussians in the source, the flux density errors are the $1\sigma$ statistical errors calculated using the formula in \citet{condon1997} and the source position is the the centroid of the Gaussians. In total 1377 sources made up of 1586 Gaussians were recovered from the image. After visual inspection, seven sources were rejected because they were found to be artefacts near bright sources. Of the remaining 1370 sources, 1251 were fit by a single Gaussian.

\subsection{Source Matching}
\label{subsec:matching}
To determine the spectral properties of the sources and check the flux density scale of our image (hereafter referred to as the VLA-P image), all of the sources were matched to the 1.4\,GHz FIRST, 1.4\,GHz National Radio Astronomy Observatory (NRAO) VLA Sky Survey \cite[NVSS;][]{nvss}, 1.38\,GHz \cite[][hereafter referred to as the deVries catalogue]{deVries2002bootes}, 325\,MHz WENSS and 153\,MHz \cite[][hereafter referred to as the Williams catalogue]{Williams2013} catalogues. A summary of the catalogues (including the VLA-P catalogue for comparison) is given in Table \ref{tbl:matching catalogues}. Matching was done with the \textsc{stilts} \citep{stilts} software package using the task \textsc{tskymatch2}. Using the central position of each sources in the VLA-P catalogue and a search radius of half the FWHM of the catalogue with the lowest resolution, \textsc{tskymatch2} returned all sources in the matched catalogue separated by less than the search radius from the input sources.

Since our aim is to search for high-redshift AGN, which are expected to have linear sizes of at most a few hundred milliarcseconds, we matched the VLA-P sources to the catalogue with the highest resolution at each frequency. However, while FIRST has the highest resolution at 1.4\,GHz, 562 sources could not be matched to a FIRST source because they were too faint to be detected in FIRST due to their steep spectral indices. Consequently, the VLA-P sources were also matched to the deVries catalogue, which, while having a lower resolution than FIRST (see Table \ref{tbl:matching catalogues}), also has a significantly lower rms noise. Hence, 417 sources that could not be matched to a FIRST source could be matched to a source in the deVries catalogue. 

\begin{table}
 \centering
 \begin{minipage}{\columnwidth}
  \caption{Catalogues matched to the VLA-P catalogue.}
  \begin{tabular}{cccc}
  \hline
  Catalogue & Frequency & Resolution & rms Noise $^{\rm a}$ \\
  Name      & [MHz]     & [$''$]     & [mJy\,beam$^{-1}$] \\
  \hline  
  FIRST    & 1400 & $5.4\times5.4$ & 0.15\\
  NVSS     & 1400 & $45\times45$   & 0.45\\
  deVries  & 1380 & $13\times27$   & 0.03\\
  VLA-P    &  325 & $5.6\times5.1$ & 0.2*\\
  WENSS    &  325 & $54\times54$   & 3.6\\
  Williams &  153 & $25\times25$   & 2.0*\\
  \hline
  \multicolumn{4}{p{\columnwidth}}{\footnotesize{\textbf{Notes:} $^{\rm a}$ The values quoted are the typical catalogue noise except for the two marked with a \lq *', which are the noise at the center of the image.}}\\
  \end{tabular}
  \label{tbl:matching catalogues}
 \end{minipage}
\end{table}

In order to exclude confused sources and sources that are resolved in one of the catalogues, only one-to-one matches were accepted. In other words, if a source in the VLA-P catalogue was matched to two or more sources in another catalogue, or vise versa, the matches were rejected. To exclude resolved sources, VLA-P sources with a major or minor axis greater than 10\,arcsec were excluded from all further analysis. While the beam size in the VLA-P image is $\sim6$\,arcsec, a value of 10\,arcsec was used to accommodate the errors on the fitted major and minor axes of the faint sources. Similarly, VLA-P sources that were matched to a source with a major or minor axis greater than 7\,arcsec in FIRST or matched to a sources that is flagged as being extended in WENSS were excluded. In total 283 resolved sources were excluded. Matching to multiple catalogues at the same frequency (FIRST, deVries and NVSS), we exclude 323 variable sources by removing sources that showed a flux density difference of more than 20\,per\,cent in any two of the catalogues. Since NVSS's resolution is about eight times lower than that of FIRST, this also ensures that none of the sources have extended structure that is resolved out in one of the higher resolution catalogues. Finally, all of the matches were visually inspected to check for potential incorrect matches. This resulted in 18 matches to 13 VLA-P sources being rejected. 

\subsection{Flux Density Scale}
\label{subsec:flux scale}
Since the central frequencies of both the VLA P-band and WENSS catalogues are 325\,MHz, we matched the two catalogues so that we could check the absolute flux density scale and the primary beam correction. When doing this, only single isolated non-extended sources were used. Additionally, sources with a signal-to-noise ratio (SNR; defined as the peak flux density divided by the island rms) below ten in either the VLA-P or WENSS catalogue were excluded from the analysis. 

\begin{figure}
  \includegraphics[width=\columnwidth]{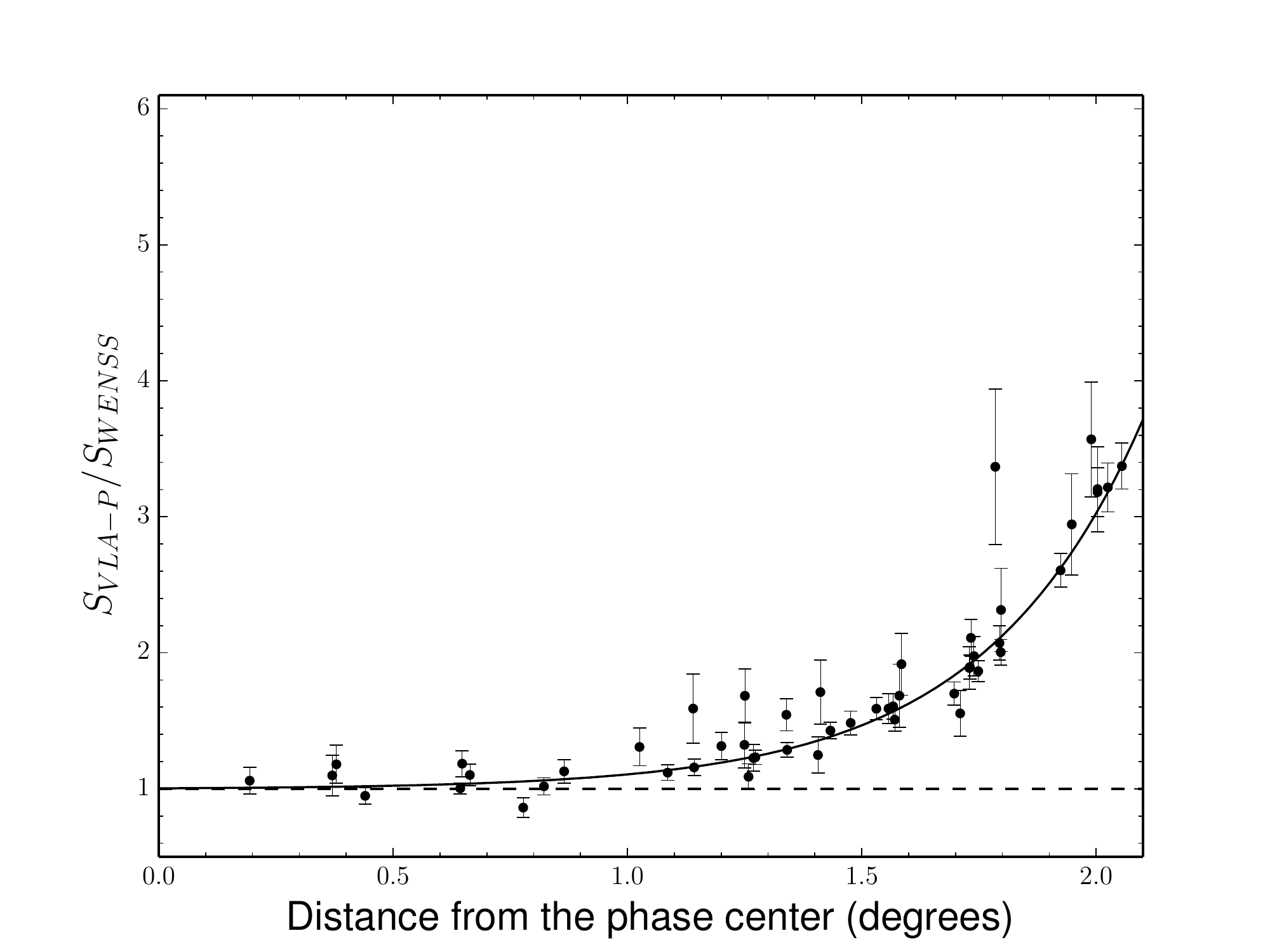}
  \caption{The uncorrected flux density ratio ($S_{\rm VLA-P}\,S^{-1}_{\rm WENSS}$) plotted as a function of the distance from the phase center for the sources used to check the flux density scale. The dashed line shows the function $y=1$ while the solid line is the function fitted to the data to correct for the VLA-P flux density offset. The function has a value of 1.0 at the phase center, increasing to 3.72 at the edge of the image.}
  \label{fig:uncorected_flux_ratio_vs_r_wenss}
\end{figure}

To check the flux density scale, the flux density ratio of the VLA-P with WENSS ($\frac{S_{\rm VLA-P}}{S_{\rm WENSS}}$) was plotted for each pair of matched sources as a function of the source SNR and the angle of the source with respect to the phase center in order to check for a potential irregular beam shape. In addition, the VLA-P flux density was also plotted as a function of the WENSS flux density for each pair of matched source. No trends were visible in any of these plots. Finally, the flux density ratio was plotted as a function of the distance of the source from the VLA-P phase center (shown in Fig.~\ref{fig:uncorected_flux_ratio_vs_r_wenss}). From the figure it is clear that sources further from the phase center (specifically those beyond the primary beam) had a larger flux density ratio, indicating that the flux densities at the edge of the VLA-P image were over estimated. This problem was likely caused by an error in the model of the VLA P-band primary beam that \textsc{casa} used to do the primary beam correction.

To correct for this offset, an exponential function of the form 
\begin{equation}
 \frac{S_{\rm VLA-P}}{S_{\rm WENSS}} = k_1 + k_2e^{k_3 r}
 \label{eq: correct flux ratio}
\end{equation}
was fitted to the data using a non-linear least squares fitting routine (shown as the solid line in Fig.~\ref{fig:uncorected_flux_ratio_vs_r_wenss}), where each point was weighted by its error. In the function, $k_1$, $k_2$ and $k_3$ are constants with best-fit values of $1.00\pm0.03$, $0.005\pm0.002$ and $2.93\pm0.19$ respectively and $r$ is the distance of the source from the phase center in degrees. The fitted function has a reduced chi-squared of 1.34, a value of 1.003 at $r = 0^\circ$ and 1.22 at $r = 1.25^\circ$, the FWHM of the VLA primary beam, increasing to 3.72 at the edge of the image. This shows that the flux density offset was only a significant problem outside the primary beam. Using Equation \ref{eq: correct flux ratio} and the condition
\begin{equation}
 \frac{S_{\rm corrected\_VLA-P}}{S_{\rm WENSS}} = k_1 ,
 \label{eq: initial condition}
\end{equation}
all of the flux densities, flux density errors and rms values, were corrected using their distance from the phase center and the errors of the fitted parameters.

\begin{figure}
  \includegraphics[width=\columnwidth]{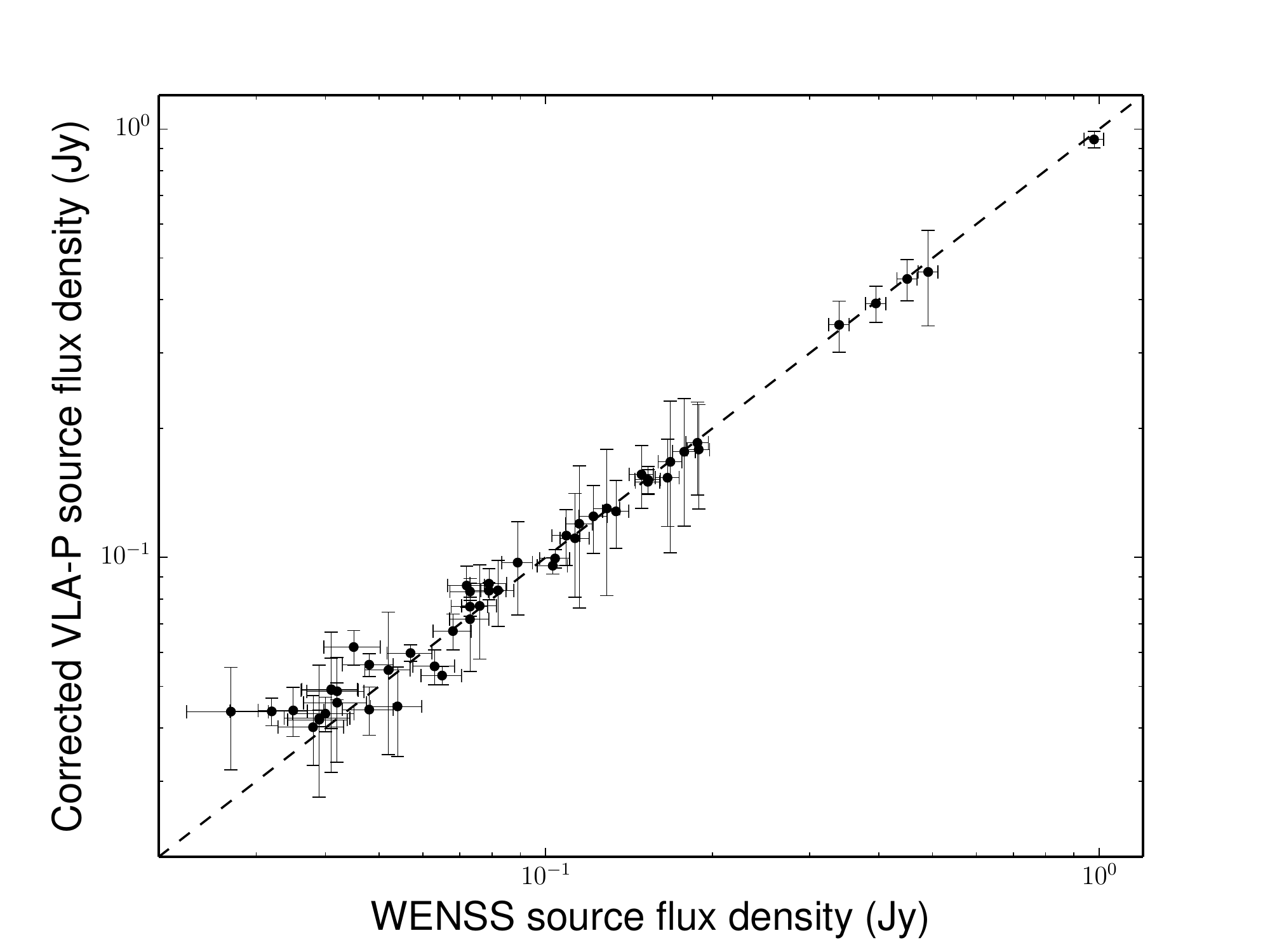}
  \caption{Log-log plot of the corrected VLA-P flux density as a function of the WENSS flux density for the sources used to check the flux density scale. The line $y=x$ was added as a guide.}
  \label{fig:cor-flux_vs_wenss-flux}
\end{figure}

Fig.~\ref{fig:cor-flux_vs_wenss-flux} shows a plot of the corrected VLA-P flux densities as a function of the WENSS flux densities for the sources used to check the flux density scale. As can be seen, the VLA-P flux densities agree well with those of WENSS. Since the mean of the corrected flux density ratios is 1.05, we conclude that no systematic offset between the VLA-P and WENSS flux densities remained after the flux density correction.

\subsection{Catalogue Construction}
\label{subsec:catalogue}
A sample of the final catalogue is given in Table \ref{tbl:catalogue_exampel}, the full catalogue containing 1370 sources is available from the CDS\footnote{http://cdsarc.u-strasbg.fr/}. In the catalogue, the flux densities, peak flux densities and rms values have been corrected as described in Section \ref{subsec:flux scale}. The columns in the table are: (1) Source name, (2,3) right ascension (RA) and uncertainty, (4,5) declination (DEC) and uncertainty, (6) integrated 342.5\,MHz flux density and uncertainty, (7) peak 342.5\,MHz intensity and uncertainty, (8,9) deconvolved major- and minor-axis, (10) position angle, (11) local rms noise, (12) number of Gaussians out of which the source is composed. In all cases, values that could not be calculated are indicated as \lq--\rq. For sources composed of more than one Gaussian, the individual Gaussians are named alphabetically and given below the source.

\begin{table*}
 \centering
 \begin{minipage}{\textwidth}
  \caption{Example entries of the online catalogue.}
  \begin{tabular}{cccccccccccc}
  \hline
  Source ID & RA    & $\sigma_{\rm RA}$ & DEC   & $\sigma_{\rm DEC}$  & $S_{i}$ & $S_{p}$ & $a$ & $b$ & $\phi$ & rms        & $N_{\rm Gauss}$\\
            & [deg] & [$''$]            & [deg] & [$''$]              & [mJy]          &[mJy\,beam$^{-1}$]& [$''$]  & [$''$]  & [deg]&[mJy\,beam$^{-1}$]& \\
  (1)       & (2)   & (3)               & (4)   & (5)                 & (6)             & (7)             & (8)     & (9)     & (10)   & (11)       & (12) \\  
  \hline  
  J143821+340235  & 219.5863 & 0.6 & 34.0432 & 0.6 &    3.6$\pm$0.9 & 2.7$\pm$0.6   & 8.2$\pm$2.0 & 4.7$\pm$0.7 & 137$\pm$18 & 0.5 & 1\\
  J143813+335310  & 219.5529 & 0.7 & 33.8861 & 0.4 &    6.6$\pm$1.0 & 3.4$\pm$0.6   & 9.1$\pm$1.6 & 6.1$\pm$0.8 & 108$\pm$21 & 0.5 & 1\\
  J143622+335939  & 219.0909 & 0.1 & 33.9942 & 0.1 & 223.5$\pm$11.0 &70.5$\pm$3.5   &14.2$\pm$0.1 & 7.1$\pm$0.1 &   9$\pm$1  & 0.4 & 3\\
  J143622+335939a & 219.0911 & 0.1 & 33.9955 & 0.1 & 107.0$\pm$5.3  &32.7$\pm$1.7   &11.5$\pm$0.1 & 8.1$\pm$0.1 &  16$\pm$2  & 0.4 & --\\
  J143622+335939b & 219.0906 & 0.1 & 33.9926 & 0.1 &  79.9$\pm$4.0  &63.0$\pm$3.1   & 6.7$\pm$0.1 & 5.4$\pm$0.1 & 117$\pm$2  & 0.4 & --\\
  J143622+335939c & 219.0908 & 0.1 & 33.9940 & 0.1 &  36.6$\pm$1.9  &22.4$\pm$1.2   & 7.8$\pm$0.1 & 6.0$\pm$0.1 & 127$\pm$4  & 0.4 & --\\
  J142208+342122  & 215.5341 & 0.7 & 34.3561 & 0.4 &  10.7$\pm$4.3  &  6.3$\pm$2.6  & 7.9$\pm$1.7 & 6.0$\pm$1.0 & 87$\pm$31  & 1.1 & 1\\
  J143948+330413  & 219.9511 & 0.8 & 33.0702 & 0.4 &  26.1$\pm$9.2  &  5.8$\pm$2.2  & 8.1$\pm$1.9 & 5.5$\pm$1.0 & 172$\pm$26 & 1.0 & 2\\
  J143948+330413a & 219.9516 & 0.7 & 33.0690 & 0.5 &   8.0$\pm$3.2  &  5.0$\pm$2.0  & 7.7$\pm$1.8 & 5.9$\pm$1.1 & 69$\pm$38  & 1.0 & --\\
  J143948+330413b & 219.9509 & 0.8 & 33.0709 & 1.0 &  18.1$\pm$6.4  &  5.3$\pm$2.1  &12.8$\pm$2.7 & 7.6$\pm$1.3 & 38$\pm$22  & 1.0 & --\\
  \hline
  \end{tabular}
  \label{tbl:catalogue_exampel}
 \end{minipage}
\end{table*}


\section{Results and Discussion}
\label{sec: results and discussion}
\subsection{Spectral Properties}
\label{subsec:Spectral Properties}
Two spectral indices, $\alpha_{low}$ and $\alpha_{high}$, were calculated for each source. The low frequency spectral index, $\alpha_{low}$, was calculated between the Williams (153\,MHz) and VLA-P (325\,MHz) catalogues. The high frequency spectral index, $\alpha_{high}$, was calculated between the VLA-P (325\,MHz) and FIRST (1400\,MHz) catalogues unless the source could not be matched to a source in FIRST (for the reason explained in Section \ref{subsec:matching}) but could be matched to a source in the deVries catalogue. In that case $\alpha_{high}$ was calculated between the flux densities of the VLA-P (325\,MHz) and deVries (1380\,MHz) catalogues.

The $\alpha_{high}$ values calculated using FIRST were found to have a median value of -0.72 and interquartile range from $-0.91$ to $-0.52$. These values are consistent with those found by \citet{smolcic2014} and \citet{owen2009} who made VLA P-band images of the VLA COSMOS and Spitzer Wide-area InfraRed Extragalactic Survey (SWIRE) fields with similar sensitivity and resolutions as our image. This serves to further confirm our flux density scale. The median value and interquartile range for $\alpha_{low}$ was found to be -0.56 and from -0.83 to -0.30. Considering that $\alpha_{high}$ and $\alpha_{low}$ have mean values of $-0.685\pm0.007$ and $-0.505\pm0.033$ respectively, is is clear that the spectral index flattens toward lower frequencies, as was reported by among others \citet{Williams2013} and \citet{vanweeren2014}. This is possibly the result of spectral aging which results in the sources having a steeper spectrum a higher frequencies.

\subsection{MPS Sources}
\label{subsec:MPS}
\begin{figure*}
  \includegraphics[width=\textwidth]{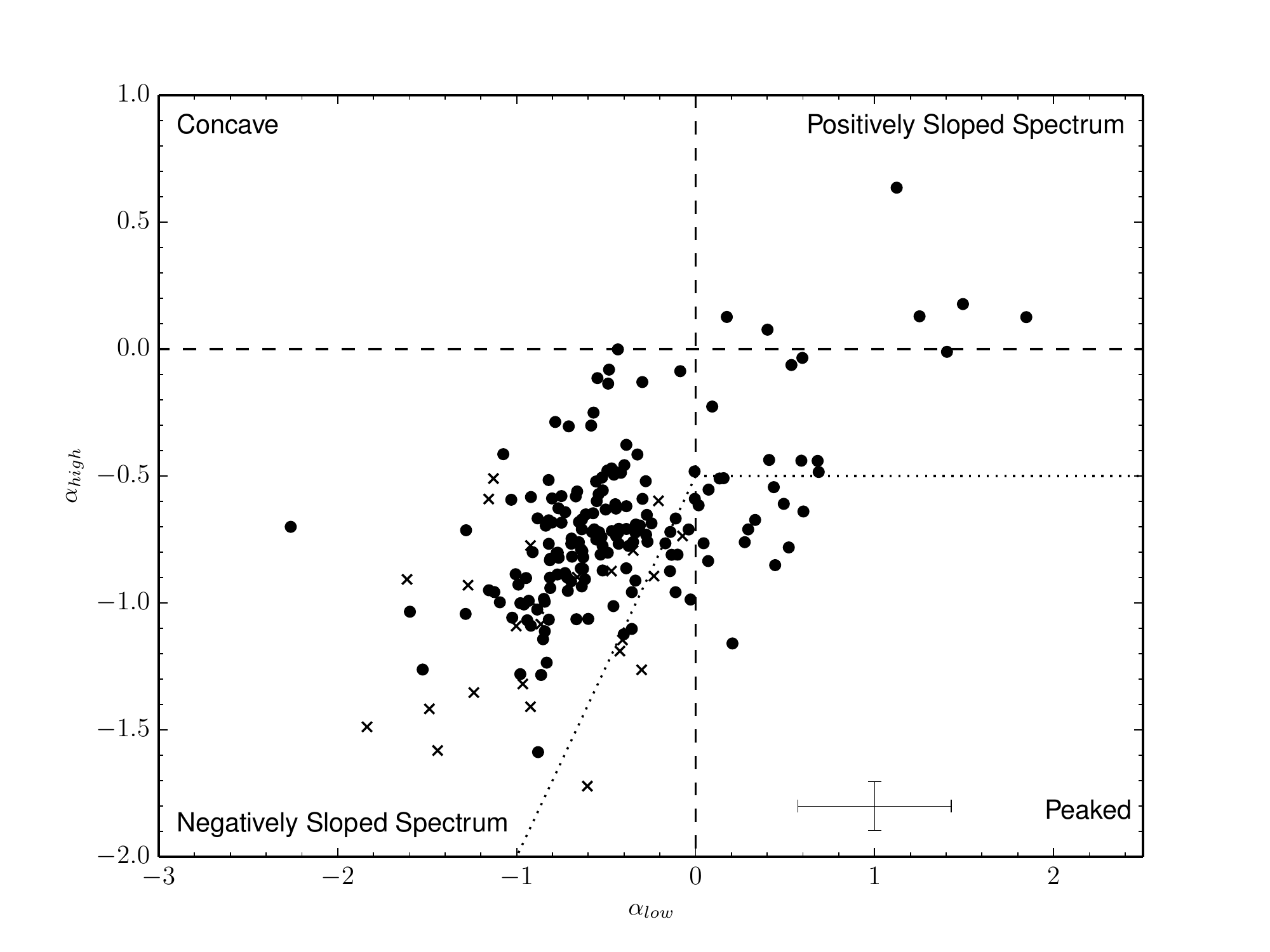}
  \caption{Colour-colour diagram for sources with a one-to-one match to the Williams and either the FIRST or the deVries catalogues. In the plot, $\alpha_{low}$ was calculated between 153\,MHz and 325\,MHz while, $\alpha_{high}$ was calculated between 342.5\,MHz and 1400\,MHz for the sources marked as filled circles and between 342.5\,MHz and 1380\,MHz for sources marked as crosses. The dashed lines indicate spectral indices of zero and the MPS sources were selected from the region below and to the right of the dotted line. While error bars are not shown to avoid confusion, the median error bar size is illustrated at the bottom right of the figure.}
  \label{fig:colour-colour}
\end{figure*}

To select the MPS sources, a colour-colour diagram (Fig.~\ref{fig:colour-colour}) was made by plotting $\alpha_{high}$ against $\alpha_{low}$ for the 198 sources which could be matched to both the Williams and either the FIRST or the deVries catalogues. In the figure, the sources in the bottom left and top right quadrants have, respectively, negative and positive spectral indices across the entire frequency range from 153\,MHz to 1.4\,GHz and contain 85 and 3\,per\,cent of the sources respectively. The remaining 12\,per\,cent of the sources lie in the bottom right quadrant and have convex (peaked) spectra. In order avoid confusion, error bars are not shown in the figure. As a guide it is noted that the median errors on the points are 0.1 and 0.43 for $\alpha_{high}$ and $\alpha_{low}$ respectively. It should be noted that from the size of the median errors, it is clear that the higher errors on $\alpha_{low}$ are caused by higher flux density errors in the Williams catalogue compared to those in FIRST.

Since the MPS sources have a spectral peak below 1\,GHz, we selected these sources to have $\alpha_{high}<-0.5$ and $\alpha_{high}<1.5\alpha_{low} - 0.5$ (the region below and to the right of the dotted line in Fig.~\ref{fig:colour-colour}). The selection criteria for $\alpha_{low}$ was used in order to not only select the sources with a clear spectral peak, but also select the sources that flatten towards lower frequencies, which could be indicative of a turnover frequency below 153\,MHz. While the selection criteria $\alpha_{high}<1.5\alpha_{low} - 0.5$ is somewhat arbitrary, it does ensure that $\alpha_{high}$ is at least one and a half times $\alpha_{low}$ for the sources that show a flattening in their spectra. In addition, this also removes the effect of the spectral index flattening towards lower frequencies (see Section \ref{subsec:Spectral Properties}) from the selection. In total, 33 MPS sources were selected and presented in Table \ref{tbl:MPS_sources}. We note that when fitting Gaussians to the sources, \textsc{pybdsm} fit a single Gaussian to each of the 33 MPS sources. In the table, the columns correspond in name to the columns in Table \ref{tbl:catalogue_exampel} except for the following columns: (8) low frequency spectral index, (9) high frequency spectral index, (10) redshift calculated using the EAZY code, (11) redshift calculated using the LTR code (a description of how columns 10 and 11 were calculated are given below). While the low frequency spectral index was calculated between 153\,MHz and 325\,MHz, the high frequency spectral index was calculated between 325\,MHz and 1400\,MHz unless the value is marked with a \lq *' in which case it was calculated between 325\,MHz and 1380\,MHz. Of the 33 sources in Table \ref{tbl:MPS_sources}, 18 have spectra that flatten ($\alpha_{low}\leq0$) while the remaining 15 sources have peaked spectra. Due to the errors on the spectral indices, only two of the 15 sources might unambigously be consistent with a peaked spectrum.

\begin{table*}
 \centering
 \begin{minipage}{\textwidth}
  \caption{List of MPS sources.}
  \begin{tabular}{cccccccccll}
  \hline
  Source ID & RA    & $\sigma_{\rm RA}$ $^{\rm a}$ & DEC   & $\sigma_{\rm DEC}$ $^{\rm a}$  & $S_{i}$ & rms        & $\alpha_{low}$ $^{\rm b}$ & $\alpha_{high}$ $^{\rm c}$ & $z_{\rm EAZY}$ $^{\rm d}$ & $z_{\rm LRT}$ $^{\rm d}$\\
            & [deg] & [$''$]            & [deg] & [$''$]              & [mJy]           & [ mJy\,beam$^{-1}$] &                      & & &\\
  (1)       & (2)   & (3)               & (4)   & (5)                 & (6)             & (7)        & (8)                           & (9) & (10) & (11)\\  
  \hline  
  J143902+351652 & 219.7583 & 0.6 & 35.2812 & 0.5 &  8.2 $\pm$ 2.2 & 0.6 & 0.21 $\pm$ 0.84  & -1.16 $\pm$ 0.19  & -- & --\\
  J143833+331845 & 219.6358 & 0.2 & 33.3126 & 0.2 & 13.9 $\pm$ 3.0 & 0.5 & 0.44 $\pm$ 0.61  & -0.54 $\pm$ 0.15  & -- & --\\
  J143813+335310 & 219.5529 & 0.7 & 33.8861 & 0.4 &  6.5 $\pm$ 1.0 & 0.5 & -0.42 $\pm$ 0.63 & -1.19 $\pm$ 0.12* & $1.793^{+0.264}_{-0.269}$ & 2.06\\
  J143657+360144 & 219.2376 & 0.5 & 36.0289 & 0.6 &  9.6 $\pm$ 3.8 & 0.9 & -0.40 $\pm$ 0.67 & -1.12 $\pm$ 0.27  & -- & --\\
  J143628+345219 & 219.1147 & 0.4 & 34.8719 & 0.5 &  4.7 $\pm$ 0.5 & 0.3 & -0.07 $\pm$ 0.55 & -0.74 $\pm$ 0.08* & $0.853^{+0.095}_{-0.094}$ & 0.77\\
  J143624+355419 & 219.0983 & 0.2 & 35.9052 & 0.5 & 12.7 $\pm$ 3.9 & 0.7 & 0.07 $\pm$ 0.60  & -0.55 $\pm$ 0.21  & -- & --\\
  J143619+350156 & 219.0805 & 0.1 & 35.0321 & 0.1 & 16.2 $\pm$ 1.3 & 0.3 & 0.29 $\pm$ 0.42  & -0.71 $\pm$ 0.07  & $0.899^{+0.051}_{-0.051}$ & 0.74\\
  J143542+330225 & 218.9244 & 0.4 & 33.0402 & 0.4 &  5.1 $\pm$ 1.0 & 0.4 & -0.41 $\pm$ 0.62 & -1.15 $\pm$ 0.14* & $2.363^{+0.161}_{-0.165}$ & 2.10\\
  J143542+341321 & 218.9259 & 0.1 & 34.2224 & 0.1 &  5.6 $\pm$ 0.4 & 0.2 & 0.44 $\pm$ 0.67  & -0.85 $\pm$ 0.06  & $2.850^{+0.106}_{-0.095}$ or & 5.30 or\\
                 &          &     &         &     &                &     &                  &                   & $2.856^{+0.108}_{-0.106}$ & 5.09\\
  J143535+325959 & 218.8958 & 0.3 & 32.9998 & 0.4 &  6.9 $\pm$ 1.2 & 0.4 & -0.03 $\pm$ 0.62 & -0.99 $\pm$ 0.12  & $3.162^{+0.191}_{-0.156}$ & 3.04\\
  J143505+334721 & 218.7707 & 0.0 & 33.7893 & 0.0 & 17.1 $\pm$ 0.9 & 0.2 & -0.14 $\pm$ 0.29 & -0.72 $\pm$ 0.05  & -- & --\\
  J143359+340421 & 218.4956 & 0.1 & 34.0725 & 0.1 &  4.2 $\pm$ 0.4 & 0.2 & -0.23 $\pm$ 0.54 & -0.89 $\pm$ 0.07* & $3.615^{+0.101}_{-0.092}$ or  & 0.62 or\\
                 &          &     &         &     &                &     &                  &                   & $1.381^{+0.174}_{-0.218}$ & 1.07\\             
  J143339+354013 & 218.4135 & 0.1 & 35.6702 & 0.2 & 10.5 $\pm$ 1.5 & 0.4 & 0.07 $\pm$ 0.44  & -0.84 $\pm$ 0.11  & $0.631^{+0.069}_{-0.071}$ & 0.54\\
  J143330+355042 & 218.3741 & 0.1 & 35.8449 & 0.2 & 14.6 $\pm$ 2.9 & 0.5 & 0.16 $\pm$ 0.48  & -0.51 $\pm$ 0.14  & $2.821^{+2.032}_{-1.536}$ & 1.37\\
  J143230+353641 & 218.1256 & 0.1 & 35.6113 & 0.3 &  9.0 $\pm$ 1.1 & 0.3 & -0.14 $\pm$ 0.43 & -0.87 $\pm$ 0.09  & $2.310^{+0.253}_{-0.248}$ & 2.36\\
  J143223+324940 & 218.0949 & 0.0 & 32.8277 & 0.1 & 27.3 $\pm$ 3.9 & 0.4 & 0.02 $\pm$ 0.35  & -0.62 $\pm$ 0.10  & -- & --\\
  J143215+350608 & 218.0621 & 0.2 & 35.1023 & 0.2 &  4.0 $\pm$ 0.4 & 0.2 & -0.30 $\pm$ 0.55 & -1.26 $\pm$ 0.08* & $0.747^{+0.043}_{-0.046}$ & 0.86\\
  J143138+353215 & 217.9064 & 0.3 & 35.5375 & 0.3 &  5.5 $\pm$ 0.7 & 0.3 & -0.11 $\pm$ 0.69 & -0.96 $\pm$ 0.10  & $0.887^{+0.091}_{-0.069}$ & 0.69\\
  J143123+331626 & 217.8472 & 0.0 & 33.2738 & 0.0 & 51.5 $\pm$ 3.1 & 0.2 & -0.13 $\pm$ 0.29 & -0.81 $\pm$ 0.05  & -- & --\\
  J143051+342614 & 217.7120 & 0.0 & 34.4372 & 0.0 & 29.0 $\pm$ 1.3 & 0.2 & 0.49 $\pm$ 0.34  & -0.61 $\pm$ 0.05  & $2.364^{+0.535}_{-0.536}$ & 2.98\\
  J143001+340746 & 217.5049 & 0.1 & 34.1295 & 0.1 &  7.4 $\pm$ 0.5 & 0.2 & 0.52 $\pm$ 0.80  & -0.78 $\pm$ 0.05  & $0.165^{+0.052}_{-0.055}$ & 0.19\\
  J142938+343903 & 217.4085 & 0.0 & 34.6507 & 0.0 & 14.5 $\pm$ 0.8 & 0.2 & -0.36 $\pm$ 0.36 & -1.10 $\pm$ 0.05  & $0.139^{+0.037}_{-0.037}$ & 0.15\\
  J142941+330552 & 217.4195 & 0.1 & 33.0978 & 0.2 &  8.6 $\pm$ 1.0 & 0.3 & 0.27 $\pm$ 0.59  & -0.76 $\pm$ 0.09  & $1.719^{+0.335}_{-0.337}$ & 1.38\\
  J142917+332626 & 217.3226 & 0.0 & 33.4407 & 0.0 & 17.3 $\pm$ 1.1 & 0.3 & 0.60 $\pm$ 0.37  & -0.64 $\pm$ 0.06  & $1.583^{+0.322}_{-0.290}$ & 2.49\\
  J142905+354425 & 217.2691 & 0.1 & 35.7402 & 0.2 & 13.0 $\pm$ 2.5 & 0.5 & -0.00 $\pm$ 0.48 & -0.59 $\pm$ 0.14  & $0.809^{+0.084}_{-0.081}$ & 0.84\\
  J142906+343326 & 217.2764 & 0.2 & 34.5571 & 0.1 &  7.9 $\pm$ 0.6 & 0.3 & -0.17 $\pm$ 0.45 & -0.77 $\pm$ 0.06  & $0.377^{+0.019}_{-0.032}$ & 0.33\\
  J142836+353154 & 217.1479 & 0.1 & 35.5315 & 0.1 & 29.7 $\pm$ 4.2 & 0.4 & -0.10 $\pm$ 0.35 & -0.81 $\pm$ 0.10  & $2.190^{+0.254}_{-0.235}$ & 2.21\\
  J142719+352324 & 216.8305 & 0.3 & 35.3900 & 0.2 &  5.1 $\pm$ 1.0 & 0.4 & -0.61 $\pm$ 0.57 & -1.72 $\pm$ 0.17* & $1.690^{+0.229}_{-0.224}$ & 1.56\\
  J142717+351756 & 216.8197 & 0.2 & 35.2989 & 0.2 &  7.7 $\pm$ 1.2 & 0.4 & 0.35 $\pm$ 0.70  & -0.67 $\pm$ 0.11  & $0.820^{+0.150}_{-0.122}$ & 0.87\\
  J142722+333123 & 216.8406 & 0.2 & 33.5230 & 0.2 &  5.2 $\pm$ 0.7 & 0.3 & 0.04 $\pm$ 0.82  & -0.76 $\pm$ 0.10  & -- & --\\
  J142608+341658 & 216.5339 & 0.3 & 34.2829 & 0.2 &  5.5 $\pm$ 0.7 & 0.3 & -0.11 $\pm$ 0.69 & -0.67 $\pm$ 0.09  & $1.329^{+0.104}_{-0.104}$ & 1.13\\
  J142602+354639 & 216.5065 & 0.5 & 35.7775 & 0.4 & 11.1 $\pm$ 4.0 & 0.8 & 0.13 $\pm$ 0.72  & -0.51 $\pm$ 0.25  & $0.859^{+0.041}_{-0.040}$ & 0.68\\
  J142310+333033 & 215.7924 & 0.4 & 33.5092 & 0.2 & 18.5 $\pm$ 6.8 & 0.9 & -0.04 $\pm$ 0.60 & -0.71 $\pm$ 0.26  & -- & --\\
  \hline
  \multicolumn{11}{p{\textwidth}}{\footnotesize{\textbf{Notes:} $^{\rm a}$ RA and DEC errors smaller than 0.04\,arcsec are rounded to 0.0. $^{\rm b}$ $\alpha_{low}$ was calculated between 153\,MHz and 325\,MHz. $^{\rm c}$ If $\alpha_{high}$ was calculated between 325\,MHz and 1380\,MHz, the value is marked with a \lq *', otherwise $\alpha_{high}$ was calculated between 342.5\,MHz and 1400\,MHz. $^{\rm d}$ The entries containing two values are because the radio source was matched to two optical sources. In the table, the redshift of both optical matches are given.}}\\
  \end{tabular}
  \label{tbl:MPS_sources}
 \end{minipage}
\end{table*}

Photometric redshifts ($z_{phot}$) were determined for the MPS sources by identifying candidate counterparts in the NOAO Deep Wide-Field Survey \citep[NDWFS;][]{ndwfs} I-band images using either the FIRST position or, if the source could not be matched to a FIRST source, the VLA-P position. We then matched the I-band sources to, and extracted flux density measurements from, the point spread function matched photometric catalogues \citep{brown2007} which comprise fluxes from the following surveys: NDWFS (B\_W, R, I and K bands), the Flamingos Extragalactic Survey \citep[FLAMEX, J and Ks bands;][]{Flamingos}, the zBootes survey \citep[z' band;][]{zBootes}, the Spitzer Deep Wide Field Survey \citep[SDWFS, 3.6, 4.5, 5.8 and 8.0 micron images;][]{sdwfs}, the Galaxy Evolution Explorer GR5 survey \citep[GALEX, near- and far-ultraviolet bands;][]{galex} and the MIPS AGN and Galaxy Evolution Survey \citep[MAGES, 24 micron image;][]{mages}. Finally, the resulting spectral energy distribution (SED) was fitted for $z_{phot}$ using the LRT code from \citet{assef2008} and EAZY code \citep{brammer2008} for comparison. A more detailed description of the matching and SED fitting of the sources in the field will be given in Williams et al. (in preparation).

In total, we were able to determine redshifts (given in Table \ref{tbl:MPS_sources}) for 24 of the 33 sources. Of the nine sources without redshifts, five lie outside the multiwavelength coverage while the remaining four were too faint to be matched. Except for the two confused sources J143359+340421 and J143542+341321, each of which have two optical matches, and J142917+332626, the values output by the LRT and EAZY codes agree well with each other and range between 0.14 and 3.16. While the LRT code, which includes an empirical AGN SED template in the fitting and therefore fits AGN spectra better, gives a mean and median redshift of 1.3 and 1.0 respectively, the mean and median redshift output by the EAZY code is 1.4 and 1.1. On average, both the LRT and EAZY codes agree that the redshift of the flattening sources are 0.1 units higher than the peaked sources. Similarly, the median redshift of the flattening sources are 0.4 units higher than that of the peaked sources for the EAZY code, and 0.3 units higher for the LRT code. While the median redshift error of the EAZY code is 0.1 units for both the flattening and peaked sources, and could therefore explain the difference between the average redshifts, this does add weight to the argument that the flattening sources could peak towards lower frequencies and could therefore be at higher redshifts than the peaked sources.

Since our current selection allows sources with an observed turnover frequency of $\sim325$\,MHz to be selected, and these sources are expected to have a linear and angular size of $\sim2.3$\,kpc and $\sim1.3$\,arcsec respectively at $z=0.1$, it is not surprising that our selection contains sources at low ($z<1$) redshifts. It should however be pointed out that these sources are classic CSS sources and hence likely to be young AGN \citep{o'dea1998,Murgia2002}. Our expectation is that the high-redshift sources should all be compact on scales of tens of milliarcseconds, which is something that we will test in our upcoming paper Coppejans et al. (in preparation), and appear as quasars at optical wavelengths \citep{o'dea1998,o'dea1990}. What is encouraging is that both codes agree that five of the sources for which we have redshifts are at $z>2$. Furthermore, given the correlation between optical magnitude and redshift for the host galaxies of sources with radio jets \citep{o'dea1998,Rocca-Volmerange2004,miley2008}, we expect that the four sources for which we could not find redshifts because they are too faint to be matched are likely at higher redshifts than the sources with known redshifts.

\subsection{USS Sources}
\label{subsec:USS}
USS source have been selected using different spectral cuts and frequency ranges eg.,  $\alpha^{\rm 608\,MHz}_{\rm 327\,MHz} < -1.1$ \citep{wieringa1992}, $\alpha^{\rm 4.85\,GHz}_{\rm 151\,MHz} < -0.981$ \citep{blundell1998}, $\alpha^{\rm 1.4\,GHz}_{\rm 843\,MHz}<-1.3$ \citep{deBreuck2004}, $\alpha^{\rm 1.4\,GHz}_{\rm 151\,MHz} < -1.0$ \citep{cruz2006}, $\alpha^{\rm 843\,MHz}_{\rm 408\,MHz} \leq -1.0$ \citep{broderick2007} and $\alpha^{\rm 1.4\,GHz}_{\rm 325\,MHz} \leq-1.0$ \citep{singh2014}. Defining USS sources as those with a spectral index steeper than -1.3, we find that only 4.2\,per\,cent of the sources for which $\alpha_{high}$ could be calculated can be classified as USS sources. Similarly 4.9\,per\,cent of the sources for which $\alpha_{low}$ could be calculated can be classified as USS sources. This is once again consistent with both \citet{smolcic2014} and \citet{owen2009} who do not find a large population of USS sources in their samples.

Looking only at the sources in the colour-colour diagram (for which both $\alpha_{high}$ and $\alpha_{low}$ are available), we find that eight sources can be classified as USS based on their value for $\alpha_{high}$ and another seven based on their value of $\alpha_{low}$. What is interesting is that only three sources have both $\alpha_{high}$ and $\alpha_{low}$ smaller than -1.3 and will appear in both selections. Hence, if we make two selections of USS sources, one based on $\alpha_{high}$ and the other based on $\alpha_{low}$, and take into account the error bars on the spectral indices, only $25^{+1}_{-8}$\,per\,cent of the sources in the combined selection appear in both of the individual selections. Changing the selection criteria for USS sources from a spectrum steeper than -1.3 to a spectrum steeper than -1.0, changes the number to $19^{+22}_{-6}$\,per\,cent. To check if these low values are because $\alpha_{low}$ is on average 0.2 units higher than $\alpha_{high}$ (see Section \ref{subsec:Spectral Properties}), we made new selections use a cut that is 0.2 units higher for $\alpha_{low}$ than that of $\alpha_{high}$. For both cases, the percentage of sources that appear in both selections were between two and five per cent lower than the values given above, indicating that the low percentages are not caused by spectral flattening at low frequencies. Hence, in our case, even when using the same spectral cut, selecting USS sources in different frequency ranges does not select the same group of sources and care should be taken when comparing different selections of USS sources.


\section{Summary and Future Prospects}
\label{sec:summary}
  
In this paper, we presented a 325\,MHz VLA image of the NOAO Bo\"{o}tes field. Both the resolution and the noise of image are an order of magnitude better than what was previously available for the field at this frequency. Matching our sources to the FIRST, \citet{deVries2002bootes} and \citet{Williams2013} catalogues of the field we calculated spectral indices for the sources between 153\,MHz and 325\,MHz and between 325\,MHz and 1.4\,GHz. We then used the spectral indices to make a radio colour-colour diagram of the unconfused point sources.

In Section \ref{sec:introduction} we described an untested method of finding young and high-redshift AGN by selecting sources with a peaked spectra in the megahertz frequency range, the megahertz peaked-spectrum (MPS) sources. Using our colour-colour diagram, we identified 33 MPS sources of which 15 have peaked and 18 have flattening spectra. Of the 33 MPS sources, we were able to determine redshift values for 24 ranging between 0.1 and 3.2. Considering that five of the sources are at $z>2$, that we expect the sources that are compact on scales of tens of milliarcseconds to be at the highest redshifts while we could only select sources that are compact on a scale of $\sim5$\,arcsec and that the four sources which were too faint to be matched are likely at $z>3$, there is encouraging evidence that the MPS method can be used to search for high-redshift sources. In the second paper in this series, Coppejans et al. (in preparation), we will take the next step in testing the MPS method by presenting observations of the sources with the European Very-long-baseline interferometry Network (EVN) to determine their angular sizes and combine that with our redshift estimates to test whether the sources at higher redshifts are compact on scales of tens of milliarcseconds as we expect.

At the moment, the greatest challenge to finding MPS sources is the lack of high resolution ($<$10\,arcsec) low noise ($<$1\,mJy\,beam$^{-1}$) radio maps below 1\,GHz. Since LOFAR will be able to make such maps in the near future and covers the frequency range between 30 and 240\,MHz, it will be ideal for searching for MPS sources. As pointed out in Section \ref{subsec:MPS}, the large errors on the low frequency spectral indices are the result of large flux density errors in the 153\,MHz Williams catalogue that was made using observations by the Giant Metrewave Radio Telescope (GMRT). Since it is expected that LOFAR will produce images with lower noise and higher resolution than what can be achieved with the GMRT at these frequencies, it will significantly improve the reliability with which MPS sources can be selected.

Finally, in Section \ref{subsec:USS} we found that selecting USS sources in different frequency ranges using the same spectral cut does not select the same group of sources and care should be taken when comparing different selections of USS sources.


\section*{Acknowledgements}

The authors wish to thank the anonymous referee, scientific editor and assistant editor for their suggestions and comments which helped to improve this paper.


\bibliographystyle{mn2e.bst}
\bibliography{references.bib}

\label{lastpage}

\end{document}